\documentclass[final,1p,times,numbers]{elsarticle}
\usepackage{array}
\usepackage{multirow}
\usepackage{rotating}
\usepackage{graphicx}
\usepackage{float}
\usepackage{supertabular} 
\usepackage{longtable}
\usepackage{amsmath}
\usepackage{fancyhdr}
\biboptions{sort&compress}

\usepackage{bm}
\usepackage{siunitx}
\usepackage{caption}
\usepackage{amsmath}
\usepackage{mathtools}
\usepackage{amsfonts}
\usepackage{graphicx}
\usepackage{xcolor}

\graphicspath{{./Pictures/}}
\journal{Progress in Aerospace Sciences}

\begin{document}
\begin{frontmatter}



\title{A Review of Gas-Surface Interaction Models for Orbital Aerodynamics Applications}


\author{Sabrina Livadiotti*\textsuperscript{a}, Nicholas H. Crisp\textsuperscript{a}, Peter C.E. Roberts\textsuperscript{a}, Stephen D. Worrall\textsuperscript{a}, Vitor T.A. Oiko\textsuperscript{a}, Steve Edmondson\textsuperscript{a}, Sarah J. Haigh\textsuperscript{a}, Claire Huyton\textsuperscript{a}, Katharine L. Smith\textsuperscript{a}, Luciana A. Sinpetru\textsuperscript{a}, Brandon E. A. Holmes\textsuperscript{a}, Jonathan Becedas\textsuperscript{b}, Rosa Mar\'ia Dom\'inguez\textsuperscript{b}, Valent\'in Ca\~nas\textsuperscript{b}, Simon Christensen\textsuperscript{c}, Anders M\o lgaard\textsuperscript{c}, Jens Nielsen\textsuperscript{c}, Morten Bisgaard\textsuperscript{c}, Yung-An Chan\textsuperscript{d}, Georg H. Herdrich\textsuperscript{d}, Francesco Romano\textsuperscript{d}, Stefanos Fasoulas\textsuperscript{d}, Constantin Traub\textsuperscript{d}, Daniel Garcia-Almi\~nana\textsuperscript{e}, Silvia Rodriguez-Donaire\textsuperscript{e}, Miquel Sureda\textsuperscript{e}, Dhiren Kataria\textsuperscript{f}, Badia Belkouchi\textsuperscript{g}, Alexis Conte\textsuperscript{g}, Jose Santiago Perez\textsuperscript{g}, Rachel Villain\textsuperscript{g} and Ron Outlaw\textsuperscript{h}}

\address{\textsuperscript{a}The University of Manchester, Oxford Road, Manchester, M13 9PL, United Kingdom\\ 
\noindent\textsuperscript{b}Elecnor Deimos Satellite Systems, Calle Francia 9, 13500 Puertollano, Spain\\
\noindent\textsuperscript{c}GomSpace AS, Langagervej 6, 9220 Aalborg East, Denmark\\
\noindent\textsuperscript{d}University of Stuttgart, Pfaffenwaldring 29, 70569 Stuttgart, Germany\\
\noindent\textsuperscript{e}UPC-BarcelonaTECH, Carrer de Colom 11, 08222 Terrassa, Barcelona, Spain\\
\noindent\textsuperscript{f}Mullard Space Science Laboratory (UCL), Holmbury St. Mary, Dorking, RH5 6NT, United Kingdom\\
\noindent\textsuperscript{g}Euroconsult, 86 Boulevard de Sébastopol, 75003 Paris, France\\
\noindent\textsuperscript{h}Christopher Newport University Engineering, Newport News, Virginia 23606, United States}

\begin{abstract}
Renewed interest in Very Low Earth Orbits (VLEO) - i.e. altitudes below $ 450 $ km - has led to an increased demand for accurate environment characterisation and aerodynamic force prediction. While the former requires knowledge of the mechanisms that drive density variations in the thermosphere, the latter also depends on the interactions between the gas-particles in the residual atmosphere and the surfaces exposed to the flow. The determination of the aerodynamic coefficients is hindered by the numerous uncertainties that characterise the physical processes occurring at the exposed surfaces. Several models have been produced over the last 60 years with the intent of combining accuracy with relatively simple implementations.
In this paper the most popular models have been selected and reviewed using as discriminating factors relevance with regards to orbital aerodynamics applications and theoretical agreement with gas-beam experimental data. More sophisticated models were neglected, since their increased accuracy is generally accompanied by a substantial increase in computation times which is likely to be unsuitable for most space engineering applications.  For the sake of clarity, a distinction was introduced between \textit{physical} and  \textit{scattering kernel theory} based gas-surface interaction models. The \textit{physical} model category comprises the Hard Cube model, the Soft Cube model and the Washboard model, while the \textit{scattering kernel} family consists of the Maxwell model, the Nocilla-Hurlbut-Sherman model and the Cercignani-Lampis-Lord model. Limits and assets of each model have been discussed with regards to the context of this paper. Wherever possible, comments have been provided to help the reader to identify possible future challenges for gas-surface interaction science with regards to orbital aerodynamic applications. 

\end{abstract}


\end{frontmatter}



Difficulties in modelling the interaction of the near-Earth aerodynamic environment with satellites in Low Earth Orbit (LEO) are due to a lack of knowledge on the mechanisms that determine
the thermosphere total density variation, the magnitude and the direction of the thermospheric wind vector and the dynamics of the Gas-Surface Interaction (GSI). These uncertainties affect the computation of the acceleration that drag - the main source of perturbation for altitudes below 600 km \citep{Lyle1971} - exerts on satellites:

\begin{equation}\label{drag}
\textbf a_{\textbf D}=-\frac{1}{2}\rho V_{rel}^{2}\frac{C_{D}S_{ref}}{m}\frac{\textbf V_{\textbf{rel}}}{|\textbf V_{\textbf{rel}}|}
\end{equation}

\noindent where $S_{ref}$ is the reference surface adopted to perform the computation
and $m$ is the satellite's mass, often the only parameter known with substantial
accuracy unless any propellant consumption needs to be acknowledged. In Eq. \ref{drag} uncertainties are found in the assessment of the total density $\left(\rho\right)$,
the aerodynamic drag coefficient $\left(C_{D}\right)$ and the satellite
velocity with regards to the rotating atmosphere $\bigl(\textbf V_{\textbf{rel}}\bigr)$. Since these sources of uncertainties are mutually linked,
any attempt to discuss them separately is improper. However, the complexity
of the problem demands some form of simplification. Therefore, challenges encountered in estimating $\rho$ and $\textbf V_{\textbf{rel}}$, whose variations are generally associated with fluctuations in the thermosphere environment, will not be treated in this review. Consequently, the key factors that for a given velocity of the flow contribute to determination of the dynamic pressure $q$ (Eq. \ref{dynamic pressure}) will be disregarded to devote more attention to those engineering variables that can be modified through proper design and materials selection:

\begin{equation}\label{dynamic pressure}
q(t) = \frac{1}{2}\rho(t) V_{rel}(t)^{2}
\end{equation}

\noindent For the reader's knowledge, comprehensive works covering the mechanisms affecting the estimate of dynamic pressure can be found in \citep{Doornbos2006, Marcos2007a, Gorney1990, Hedin1987,Danilov2001, Hunsucker1982, Richmond1978}. 

In the following sections of this paper, the effect of GSI dynamics on drag evaluation and computation of the aerodynamic coefficients will be discussed. Some information regarding the aerodynamic regime experienced by satellites in LEO - and especially in Very Low Earth Orbit (VLEO) - will be provided in an attempt to create an adequate background for the discussion. Attention will be focused on statistical and physical GSI models which have obtained considerable success both in engineering applications and surface science for their capability to describe the complex processes occurring at the surface with relative simplicity.  In this regard, the intention is to create continuity with a recent review paper on the topic by \citet{MostazaPrieto2014}, which focuses on classical analytical models used for aerodynamic computation in LEO. Gas-beam experimental results conducted in the physical regimes of interest for this paper will finally be presented, and the behaviour of the models described will be discussed, wherever possible, against the identified trends in scattering. The objective is thus to highlight the points of strength and the limits of the theoretical models against the available experimental data. In this way possible feature developments can be discussed and hopefully a reasonable picture of the difficulties encountered in approaching the GSI problem for orbital aerodynamics can be provided.
 
Increased knowledge of the interaction mechanisms occurring in the gas-solid phase system is crucial not only for scientific achievement, but also for the possibility of improving aerodynamic performance of spacecraft operating in VLEO \citep{Roberts2017,Crisp2020}. This would reflect in increased confidence in assessing the advantages and the drawbacks of employing aerodynamic torques for orbit \citep{Bevilacqua2008, Perez2012, Dellelce2015, VirgiliLlop2015, Virgili2015, Leppinen2016, Omar2018} and attitude control purposes \citep{Pande1979,Stengle1980,Gargasz2007, Auret2011,Virgili2013,Llop2014,Hao2016,MP2017,Mostaza-Prieto2017, Canas2018}. Overestimating or underestimating the aerodynamic torques induced by the actuation of aerodynamic control surfaces has an impact on the altitude range for which aerodynamic manoeuvring is expected to be feasible. This seems relevant especially for missions operating in periods of low solar activity, when the thermospheric density values at altitudes above 200 km are significantly lower than those expected during high solar activity \citep{MostazaPrieto2014}. In terms of attitude control implementation, the achievable aerodynamic control authority about the roll, pitch and yaw body axes drives the design of the controller selected. This is especially true if conventional actuators (e.g. reaction wheels, magnetorquers) are meant to be used in synergy with the designated control surfaces. Undesired aerodynamic torques may counteract the control action of the wheels, disturbing the attitude task and eventually leading to saturation. Similarly, aerodynamic orbit control \citep{Palmerini2014}, formation flying \citep{Lambert2012, Shao2017, Spiller2017, Sun2017} and rendezvous manoeuvres \citep{Horsley2013, Smith2017} would significantly benefit from any improvement in GSI models, especially with regards to the possibility of producing control torques in the direction perpendicular to the orbit plane. A reliable estimation of the aerodynamic coefficients is also fundamental in the evaluation of the impact that aerodynamic based manoeuvres may have on the rate of decay of satellites in orbit. This knowledge could potentially be useful for a better prediction of the satellite re-entry trajectory \citep{Virgili2015,Omar2018} or to achieve a better knowledge of the expected aerodynamic forces and torques induced on the ram surfaces during controlled  re-entry in atmosphere. This knowledge represents a considerable advantage even for spacecraft that are already in orbit, assuming that the materials employed for the external surfaces are known and that a good prediction of the environmental conditions is achiavable. Nevertheless, a better knowledge of the phenomena occuring at the surface could potentially drive a more rational design of the satellite geometry according to the desired aerodynamic performances. The same geometry is indeed expected to provide a different aerodynamic behaviour with varying scattering re-emission patterns. Comparably, the design of Atmospheric Breathing Electric Propulsion (ABEP) systems is driven by the aerodynamic performance expected for the materials employed \citep{Romano2015}. Any improvements in the reliability of the GSI models may translate in a more confident definition of optimal performance ranges and it can possibly pave the way for new design criteria.

\section{The Aerodynamic Environment}
At the altitudes where VLEO satellites orbit, i.e. below 450 km \citep{Virgili-Llop2014},  the atmosphere is so
tenuous that the flow can no longer be considered a continuum. In
this scenario, the principles that rule the interaction of gas constituents
with each other and with a body immersed in the flow change, as
do the assumptions used to investigate the aerodynamic environment.
The discriminating factor employed to distinguish between the different regimes
is a dimensionless ratio known as the Knudsen number $\left(K_{n}\right)$,
which compares the order of magnitude of the molecular mean free path
$\left(\lambda\right)$ with a characteristic dimension of the flow field 
$\left(L\right)$:

\begin{equation}
K_{n}=\frac{\lambda}{L}
\end{equation}
\label{Aerodynamic env}

As evidenced by the definition, the Knudsen number is not strictly
a flow property since its value is partially controlled by the reference
length adopted to describe the field of motion. $L$ is generally
assumed to coincide with a significant dimension of a body in the flow, but
this association is not unique. Between $\lambda$ and $\rho$
there is an inverse proportional correlation, according to which,
high values of the mean free path (and thus $K_{n}$) are usually
associated with low density levels or gas-surface interactions occurring at nano-scale length. 

Three fundamentals regimes are usually identified according to the
Knudsen number. Small $K_{n}$ $\left(0<K_{n}<0.1\right)$
are typical of the familiar continuum dynamics, where collisions between
particles is the prevalent mechanism of interaction. When $K_{n}\rightarrow\infty$,
the length travelled by the particles before impingement with other particles in the gas mixture is considerable
compared to the characteristic flow-field dimension. In these conditions, the flow is characterised by a structure in which the interaction of the gas-particles
with the surface dominates over inter-particle collisions. The
distance a reflected particle travels before colliding with the
free stream is of the same order of magnitude of $\lambda$ and consequently, the flow itself can be considered collisionless. In these conditions,
it can be assumed that the presence of a body in the flow-field does
not perturb the distribution of the incident stream of particles in
the vicinity of the body itself, and consequently, no shock waves
are expected to arise. This behaviour is typical of a highly rarefied
flow regime, more commonly known as \textit{free molecular flow} (FMF).
The majority of authors agree to set $K_{n}>10$ as the lower limit
to identify this region \citep{Mostaza-Prieto2017,MostazaPrieto2014,VirgiliLlop2014,Sutton2009,Sentman1961,Storch2002,Bird1994}, with some variations \citep{Schaaf1958}. Intermediate $K_{n}$
values identify the near-free molecular regime $\left(K_{n}\gg1\right)$
and the complex transitional regime $\left(K_{n}\sim1\right)$, where
continuum and rarefied dynamics phenomena are of similar relevance
\citep{Ivchenko2007}. 

The determination of the Knudsen number for the orbital aerodynamics problem is especially dependent on a judicious choice of the mean free path for the coordinate system used. For the free molecular assumption to be valid, 1) the interactions of the incident particles with each other and 2) the interactions of the incident particles with the reflected particles should be negligible compared to the probability of collision of the incident particles with the surface. For conditions 1) and 2) to be met simultaneously,  it is necessary that the order of magnitude of the associated mean free paths is such that:

\begin{equation}
K_{n, \lambda_{ii}} =\frac{\lambda_{ii}}{L} > 10 \;\; \;\;\;\; \;\; K_{n, \lambda_{ir}} =\frac{\lambda_{ir}}{L} > 10
\end{equation}

\noindent where $ \lambda_{ii}$ is the mean free path referring to the interaction with other incident particles and $\lambda_{ir}$ is the mean free path related to the interaction with the reflected particles. According to its definition in kinetic theory, the mean free path  is inversely proportional to the effective collision cross-sectional area ($\pi d^{2}$) and the number density of the particles ($n$) \citep{Sentman1961}:

\begin{equation}
\lambda\propto\frac{1}{\pi d^{2}n}
\end{equation}

\noindent where $d$ is the radius of the sphere of influence. For the problem examined, the velocity of the satellite through the gas is considerable and the re-emission of the particles from the surface is typically thermal. In these conditions, the number density of the reflected molecules can significantly increase, especially in proximity of the surface, possibly changing the nature of the flow locally. If the coordinate system is assumed to be fixed with the body immersed in the flow, $\lambda_{ir}$ tends to be an order of magnitude smaller than $\lambda_{ii}$ \citep{Sentman1961} and should thus be preferred for a conservative estimation of $K_{n}$. For many applications, however, the free stream mean free path ($\lambda_{\infty}$), defined with regards to a coordinate system fixed with the gas, is adopted:

\begin{equation}
\lambda_{\infty} = \frac{1}{\sqrt{2} \pi d^{2} n_{i}}
\end{equation}

\noindent where $n_{i}$ is the number density of the free stream incident particles. The relation between $\lambda_{ir}$ and $\lambda_{\infty}$ is provided by \citet{Sentman1961}:

\begin{equation}
\lambda_{ir} = \sqrt{\frac{\pi}{2}}\frac{1}{s}\sqrt{\frac{T_{r}}{T_{i}}}\lambda_{\infty}
\end{equation}

\noindent where $s$ is the molecular speed ratio defined later in this section, $T_{r}$ is the temperature of the reflected particles and $T_{i}$ is the temperature of the incident particles. Generally for satellites in VLEO, $s >5$ and $T_{r}/T_{i} < 1$. As a consequence, $\lambda_{\infty}$ can be considerably bigger that $\lambda_{ir}$. Results shown by Walker et al. \citep{Walker2014} suggest that the $K_{n}$ number referred to freestream should be in the order of $10^{3}$ to assure overall free molecular conditions. Detailed analysis of the uncertainties related to the $K_{n}$ number computation for orbital aerodynamics applications is provided in \citep{Sentman1961}. 

Because of the extremely low density of the upper atmosphere, VLEO is generally characterised as a FMF environment. While the Knudsen number defines
the degree of rarefaction of the gas, another parameter is needed
to indicate how the relative magnitude of the satellite's velocity
and the most probable gas velocity affects the aerodynamics. Just
as the Mach number expresses the relationship between the body's macroscopic
velocity and the speed of sound, the molecular speed ratio indicates
the ratio between the gas macroscopic velocity $\left(v_{m}\right)$ and the most probable molecular
thermal velocity $\left(v_{t}\right)$ according to a Maxwell-Boltzmann distribution:

\begin{equation} \label{molecular speed ratio}
s=\frac{v_{m}}{v_{t}}=\frac{V_{\infty}}{\sqrt{\frac{2RT_{\infty}}{m_{m}}}}
\end{equation}

\noindent In Eq. \ref{molecular speed ratio}, $R$ is the gas constant, $m_{m}$ is the gas molecular mass and $T_{\infty}$
is the gas kinetic temperature of the free stream. For the orbital aerodynamics problem, the velocity of the body immersed in the FMF is so high that the investigation concerns the motion of a body travelling at high speeds through a gas at rest. Consequently, in Eq. \ref{molecular speed ratio} the gas macroscopic velocity effectively corresponds to the satellite's velocity $\left(V_{\infty}\right)$ and the two can be used interchangeably. 

At a certain altitude, variations in thermal velocity - and thus internal energy - occur with alterations in the amount of energy absorbed by the atmosphere \citep{Jacchia1977, Gorney1990,Hedin1987a}.
In these conditions, the random thermal velocity may play a role in the determination of the induced aerodynamic forces. This behaviour is generally associated with small values of $s$ and the FMFs are accordingly referred as \textit{hypothermal} flows.


 On the contrary, when $s\rightarrow\infty$, the effect of the bulk
velocity of the particles on the aerodynamic forces estimation is considered predominant. Under these conditions, the flow is said to be \textit{hyperthermal} and approximate kinetic theories ignoring the drift caused by the random thermal motion of the particles are usually preferred. Typical values
of $s$ at VLEO altitudes are greater than 5. For general applications, the hyperthermal
assumption is considered valid for $s>5$ \citep{Mostaza-Prieto2017,MostazaPrieto2014}
and, for this reason, it is frequently employed. 

The particulate flow impinges on the external surfaces of a spacecraft, generating
induced aerodynamic forces whose magnitude is only dependent on the nature
of the interaction. The aerodynamic forces experienced by the satellite in the body-axes reference system conventionally used for flight-mechanics applications can be modelled referring to the familiar expression:

\begin{equation}\label{aerodynamic forces}
\textbf F_{\textbf{aero}}=m\textbf a=\frac{1}{2}\rho\ V_{rel}^{2}S_{ref}\textbf C_{\textbf{F}}
\end{equation}

\noindent where, similarly to Eq.\ref{drag}, $\rho$ is the thermospheric density, $V_{rel}$ is the satellite velocity with regards to the oncoming flow, $S_{ref}$ is the selected reference surface and $\mathbf{C_{ F}}=\left[C_{A}, C_{S},C_{N}\right]^{T}$
is the vector of the three aerodynamic components along the axial, the side and the normal direction, respectively. 
Similarly, the resulting
aerodynamic torques referenced to the centre of mass are given
by: 

\begin{equation}\label{aerodynamic torques}
\textbf T_{\textbf{aero}}=\textbf r_{\textbf{PO}}\times m \textbf a=\frac{1}{2}\rho V_{rel}^{2}S_{ref}l_{ref}\textbf C_{\textbf M}
\end{equation}

\noindent where $\mathbf{C_{M}}=\left[C_{l},C_{m},C_{n}\right]^{T}$ is the vector of
the roll, pitch and yaw momentum coefficients and $\textbf r_{\textbf{PO}}$
is the position vector defining the distance between the aerodynamic centre of
pressure and the centre of mass. The magnitude of the aerodynamic
forces and coefficients has been estimated in literature for bodies
of different shapes making use of both analytical \citep{Fuller2009, Hart2015,Hart2014,Gaposchkin2008,Storch2002} and numerical techniques \citep{Jin2016, Mehta2014}. Both approaches have benefits and drawbacks
and, ideally, the most advantageous strategy would be to adopt them
in synergy, when permitted. 

Regardless of the simulation technique, the estimation of the aerodynamic coefficients relies on the models employed to physically describe the
underlying mechanism for GSI. $\mathbf{C_{F}}$ and $\mathbf{C_{M}}$
are generally computed extending the integrals of the local stress coefficients
$\left(c_{F}\right)$ to the surface exposed to the incident flow:

\begin{equation}\label{CF}
\mathbf{C_{F}}=\int_{S}\mathbf{c_{F}}\:dS
\end{equation}

\begin{equation}\label{CM}
\mathbf{C_{M}}=\int_{S}\mathbf{r_{PO}}\times\mathbf{c_{F}}\:dS=\int_{S}\bigl(\mathbf{r_{P}}-\mathbf{r_{O}}\bigr)\times\mathbf{c_{F}}\: dS
\end{equation}

\begin{figure}
\centering
\includegraphics[scale=1]{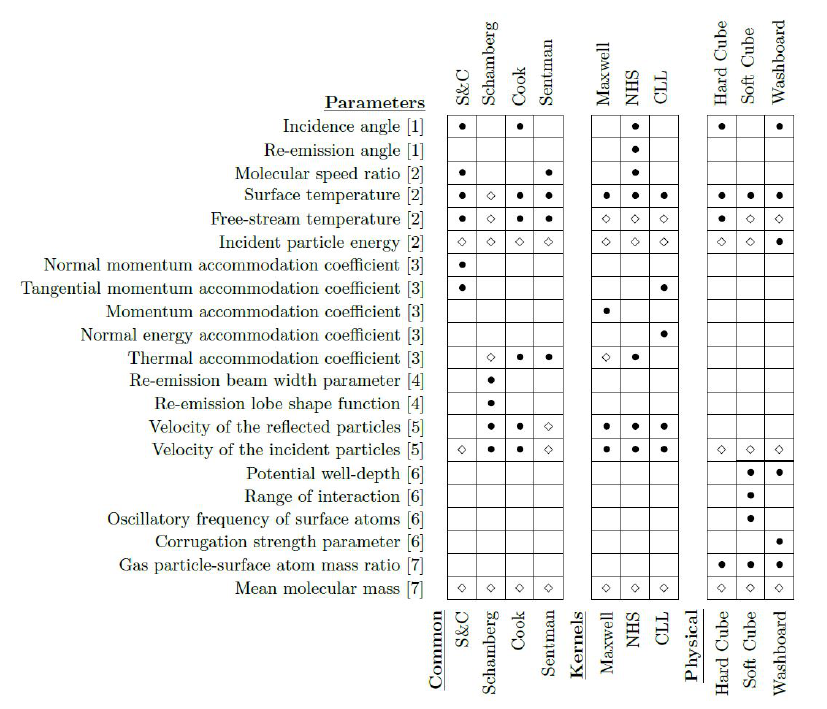}
\caption{GSI models compared: parameters of dependence for scattering distributions and aerodynamic coefficients. Black dots are used to identify parameters explicitly declared in the formulations, while white rhombus indicate implicit parameters of dependence. Parameters are grouped according to the following families: [1] flow angles; [2] energy parameters; [3] accommodation coefficients; [4] beam shape parameters; [5] velocities; [6] interaction parameters; [7] mass parameters.}
\label{model_param}
\end{figure}

\begin{figure*}
\centering
\includegraphics[scale=1]{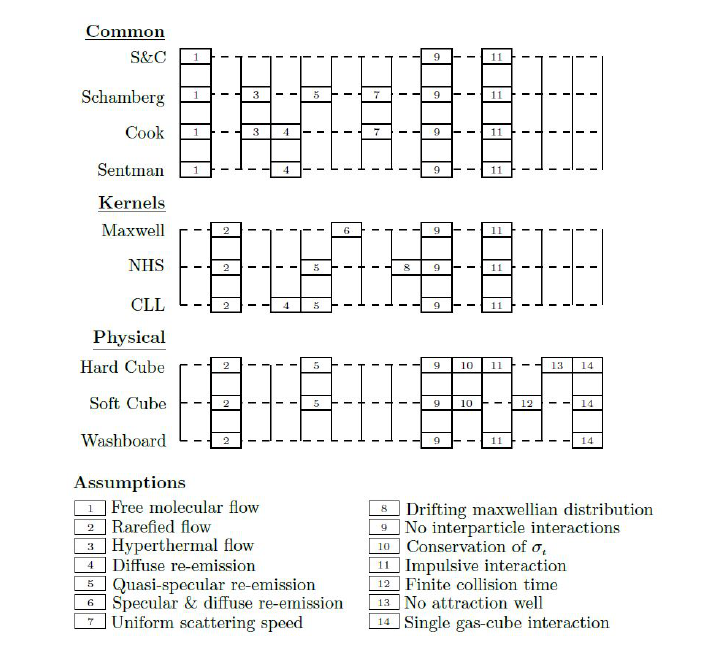}
\caption{Comparison between GSI models assumptions.}
\label{model_assumptions}
\end{figure*}

\noindent The variety of proposed GSI models translates into a variety in $c_{F}$
formulations. In particular, the expressions found in literature suggest
that the aerodynamic coefficients are a complex function of a number
of parameters which, once again, vary with the model adopted
(Fig. \ref{model_param} and \ref{model_assumptions}). A certain agreement is however found in the use of the so-called accommodation
coefficients, the wall temperature which is usually assumed constant,
the incident gas kinetic temperature and the molecular speed ratio
$s$. These parameters are likely to depart from their initial value with variations in surface contamination,
composition and structure, surface thermal properties and incident
particle energy and velocity. Further dependencies on geometry and velocity
vector direction are incorporated by the components of stress acting perpendicularly and tangentially to the surface. 


The incident free stream, assumed to be in local Maxwellian equilibrium, interacts with
the surface transferring both energy and momentum to it. Kinetic theory-based GSI models try to describe the physics underlying this phenomenon according to the contribution coming from both the incident and re-emitted stream of particles, with major difficulties found in establishing a satisfactory mathematical model for the latter.
The amount of energy and momentum exchanged is a measure of the equilibrium the
impinging particles achieve with the surface before re-emission. Both
phenomena are described by means of a set of average phenomenological coefficients.
The \textit{thermal} or \textit{energy accommodation coefficient},
first introduced by Knudsen \citep{Knudsen1911}:

\begin{equation}
\alpha_{T}=\frac{E_{i}-E_{r}}{E_{i}-E_{w}}=\frac{T_{k,i}-T_{k,r}}{T_{k,i}-T_{w}}
\label{accommodation coeff}
\end{equation}

\noindent describes the energy exchange, assuming
that the translational, rotational and vibrational energies of the
particles are all affected to the same degree by the interaction with
the wall \citep{Schaaf1958}. In Eq. \ref{accommodation coeff}, $E_{i}$ and $E_{r}$
are the kinetic energies carried by the incident and the scattered
fluxes, while $E_{w}$ denotes the energy that would be carried away
from the surface by the scattered flux if complete thermal equilibrium
was achieved and particles were re-emitted according to a Maxwellian
distribution corresponding to the surface temperature $\left(T_{w}\right)$.
Similarly, $T_{k,i}$ and $T_{k,r}$ indicate the kinetic temperatures
of the incident and reflected streams. In accordance to what will be discussed in the following sections, it is also appropriate to introduce a \textit{partial thermal accommodation coefficient} \citep{Hurlbut1968}, whose value depends on the specific incident ($\theta_{i}$) and scattering ($\theta_{r}$) directions selected with regards to the normal to the surface:

\begin{equation}\label{partial thermal accommodation}
\alpha_{T,P}\bigl(\theta_{i},\theta_{r}\bigr)=\frac{E_{i}\bigl(\theta_{i}\bigr)-E_{r}\bigl(\theta_{r}\bigr)}{E_{i}\bigl(\theta_{i}\bigr)-E_{w}}
\end{equation}

To describe the momentum exchange,
it is common practice to refer to the momentum coefficient $\left(\sigma\right)$  \citep{Maxwell1890}.
 Better physical correlation is usually achieved by adopting two separate accommodation
coefficients to describe the normal ($\sigma_{n}$) and the tangential
($\sigma_{t}$) momentum exchange \citep{Schaaf1958}:

\begin{equation}\label{sigman}
\sigma_{n}=\frac{p_{i}-p_{r}}{p_{i}-p_{w}}
\end{equation}

\begin{equation}\label{sigmat}
\sigma_{t}=\frac{\tau_{i}-\tau_{r}}{\tau_{i}}\qquad\tau_{w}=0
\end{equation}

\begin{figure*}
\centering
\def\svgwidth{120mm}
\includegraphics[width = 120mm]{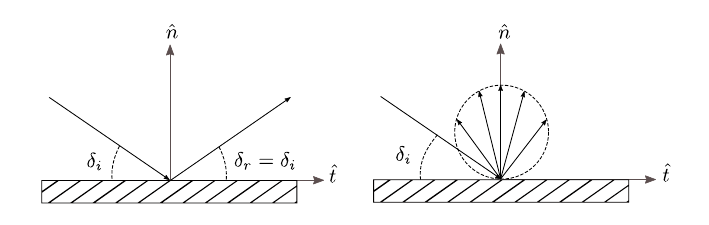}
\caption{Mechanisms of re-emission for specular reflection without thermal accommodation (left) and diffuse reflection with complete thermal accommodation (right).}
\label{SpecularDiffuse}
\end{figure*}

\noindent The quantities in Eq. \ref{sigman} and \ref{sigmat} are analogous to those already described
for $\alpha_{T}$, the only difference being that in this case they
refer to the momentum rather than the energy of the fluxes. The
information of most significant value provided by $\alpha_{T}$, $\sigma_{n}$
and $\sigma_{t}$ is that the distribution of the re-emitted particles and
velocity is deeply influenced by the degree of accommodation of the
incident molecules with the surface. By referring to these quantities,
two classical and extreme mechanisms of interaction are identified, namely specular reflection and diffuse re-emission. If specular
reflection occurs without any thermal accommodation, the molecules
are elastically reflected, no thermal energy is transferred
to the body and momentum exchange occurs only along the normal to the surface $\left(\alpha_{T}=\sigma_{n}=\sigma_{t}=0\right)$. The angle
that the velocity vector of the reflected particles form with the
surface is equal to the one of the incident stream and it lies
in the same plane of the incident velocity vector and the normal to
the surface (Fig.\ref{SpecularDiffuse}, left).

\noindent If isothermal diffuse re-emission with complete thermal
accommodation occurs ($\alpha_{T}=\sigma_{n}$ $=\sigma_{t}=1$),
particles have time to reach equilibrium with the surface and they
are re-emitted according to a probabilistic velocity and direction
distribution determined by the wall temperature, regardless of the
incident stream's history (Fig.\ref{SpecularDiffuse}, right). Experimental results discussed in Section \ref{gas_beam} however suggest that more complex scattering interactions
are expected to occur at the surface.

In VLEO, solar emissions in the Extremely Ultra Violet (EUV) wavelength have sufficient energy to generate Atomic Oxygen (AO) from the diassociation of $O_{2}$. 
The chances for the high reactive AO to reassociate to form $O_{2}$ or $O_{3}$ are quite low because of the very large mean free path characterising extremely rarefied flow regimes.
Because of this, AO represents the main atmospheric constituent at VLEO altitudes and the main source of contamination and degradation of surfaces exposed to the flow. Effects of AO interaction with polymers and metals include oxygen erosion and inclusion, as well as formation of volatile and non-volatile reaction products \citep{Banks2004}.
Because of the high degree of contamination of the surfaces in VLEO,
most works assume diffuse reflection with complete thermal accommodation.
However, if highly accommodated particles are likely to be predominant
below 300 km \citep{Moe2005}, the same cannot be said at higher altitudes,
where the atmosphere gradually
becomes less and less dense, thus limiting the contaminant adsorption to the spacecraft surfaces.
%
%
According to this, Moe and Moe \citep{Moe2005} proposed a Maxwellian-like
model to compute the drag coefficient. The latter uses a modified form of
Sentman's model to compute the contribution associated with the $0<\sigma<1$
fraction of particles which are diffusely re-emitted. Schamberg's model \citep{Schamberg1959} is
instead used in combination with Goodman's accommodation coefficient \citep{Goodman1967a} to address the input coming from the $\left(1-\sigma\right)$
fraction that is quasi-specularly reflected. Schamberg's \citep{Schamberg1959} and
Sentman's \citep{Sentman1961} are probably the most popular models
adopted to perform the estimation of aerodynamic properties. However, like
many other GSI models, they rely on a specific set of assumptions
that restrict their range of applicability. Schamberg's quasi-specular
model assumes hyperthermal impinging FMF and uniform scattering speed
along all directions. The hyperthermal approximation, conserved by Cook in a re-adaption of the model \citep{Cook1966}, provides valid
results for applications in VLEO. However some care must be taken
since, as more rigorously discussed in \citep{Sentman1961}, neglecting the particle thermal velocity may introduce some errors for small angles of attack even for $s>5$ . 

\begin{figure} 
\begin{centering}
\def\svgwidth{120mm}
\includegraphics[scale=1]{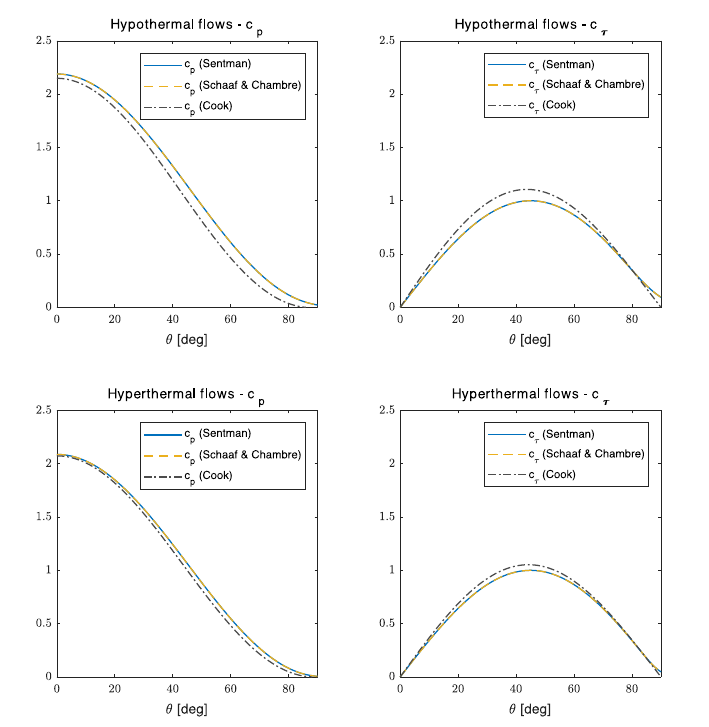}
\par\end{centering}
\caption{Comparison between Sentman's \citep{Sentman1961}, Cook's \citep{Cook1966} and Schaaf \& Chambre's \citep{Schaaf1958} model for hypothermal flows ($s = 6$) and hyperthermal flows ($s = 12$). The values of $ c_{p}$  and $c_{\tau}$ predicted by the models are shown, assuming $\alpha_{T}$ = 1,  $\sigma_{n} = 1$,  $\sigma_{t} = 1$,  $T_{w}$ = 300 K  and $T_{\infty}$= 1000 K and a flat plate as reference geometry. }
\label{model comparison}
\end{figure}

The effect of neglecting particle thermal velocity is more visible when the aerodynamic coefficients
are expressed in terms of normal $\left(c_{p}\right)$ and shear $\left(c_{\tau}\right)$ stress components:

\begin{equation}
C_{D}\approx c_{p}cos\theta+c_{\tau}sin\theta
\end{equation}

\begin{equation}
C_{L}\approx c_{p}sin\theta-c_{\tau}cos\theta
\end{equation}

\noindent Fig. \ref{model comparison} shows the aerodynamic behaviour of a flat plate immersed in free molecular flow, according to a hyperthermal model (Cook \citep{Cook1966}) and non-hyperthermal models (Schaaf \& Chambre \citep{Schaaf1958}, Sentman \citep{Sentman1961}). For varying molecular speed ratios, variations of $c_{\tau}$ are noticeable for $\theta\sim90^{\circ}$, and thus grazing angles of attack. The predicted $c_{p}$ values vary as well, showing noticeable deviation and disagreement between hypothermal and hyperthermal models at low $s$ values.  
On the other hand, Sentman's model considers molecules random thermal motion, but it is built on the assumption of complete diffuse reflection.
These models will not be discussed in detail here, since a comprehensive review can be found in \citep{Mostaza-Prieto2017, MostazaPrieto2014}. However, their fundamental assumptions are summarised in Fig. \ref{model_param} and \ref{model_assumptions}.

\section{Scattering-kernel theory based GSI models}\label{Section:Phenomenological GSI models}
 Scattering-kernel models, as we will refer to them in this section, are kinetic models built on a statistical approach. Correlation between the incident and reflected distributions of the particles is achieved through a proper definition of the boundary conditions for the collisionless Boltzmann equation in the rarefied gas regime. The correct formulation of these is of fundamental importance since boundary conditions describe the mechanisms that rule the interaction between gas and solid particles. When boundary conditions are written within the frame of the scattering-kernel theory, the problem consists in finding the most suitable expression for the scattering kernel $K\left(\mathbf{x}, t, \bm{\xi_{i}}\rightarrow\bm{\xi_{r}}\right)$ to reproduce the interaction phenomena occuring at the surface under the assumptions considered.

 Since an accurate knowledge of the dynamics and the thermodynamics of the interaction is not achievable, a quantity $P\left(\bm{\xi_{i}}\rightarrow\bm{\xi_{r}}, \mathbf{x_{i}}\rightarrow \mathbf{ x_{r}}, t_{i}\rightarrow t_{r}\right)$ can be introduced to describe the density of probability that a gas particle impacting on a point $ \mathbf{x_{i}}$ located on the surface, at a certain time $t_{i}$  with an incident velocity corresponding to $\bm{\xi_{i}}$ is reflected at a generally different location $\mathbf{ x_{r}}$ at time $t_{r}$ with a velocity $\bm{\xi_{r}}\neq\bm{\xi_{i}}$ \citep{Cercignani1971}. Assuming that  the sitting time on the surface is low enough ($t_{i}\simeq t_{r}=t \rightarrow \mathbf{x_{i}}\simeq \mathbf{x_{r}}=\mathbf{ x}$) such that no adsorption/desorption or diffusion phenomena need to be addressed \citep{Cercignani1971}:

\begin{equation}
P\left(\bm{\xi_{i}}\rightarrow\bm{\xi_{r}}, \mathbf{x_{i}}\rightarrow\mathbf{x_{r}}, t_{i}\rightarrow t_{r}\right)= K\left(\bm{\xi_{i}}\rightarrow\bm{\xi_{r}}\right)
\end{equation}

\noindent where the incident and reflected molecular velocity vectors more generally consist of two tangential ($\bigl[\xi_{i,t'},\xi_{i,t''}\bigr]$, $\bigl[\xi_{r,t'},\xi_{r,t''}\bigr]$) and one normal ($\xi_{i,n}<0$, $\xi_{r,n}>0$) component. 
%
Proposed classes of $K\left(\bm{\xi_{i}}\rightarrow\bm{\xi_{r}}\right)$ not only need to correctly capture the phenomena occuring in the proximity of the wall through a proper correlation of the incident  $\bigl(f_{i}\bigl(\bm{\xi_{i}}\bigr)\bigr)$ and reflected $\bigl(f_{r}\bigl(\bm{\xi_{r}}\bigr)\bigr)$ velocity distributions:

\begin{equation} \label{balance_at_the_surface}
\xi_{r,n}f_{r}\bigl(\bm{\xi_{r}}\bigr)=\int_{\xi_{i,n}<0}K\bigl(\bm{\xi_{i}}\rightarrow\bm{\xi_{r}}\bigr)\lvert\xi_{i,n}\rvert f_{i}\bigl(\bm{\xi_{i}}\bigr)d\bm{\xi_{i}}
\end{equation}

 \noindent but also need to satisfy some specific conditions. The non-negativity condition \citep{Cercignani2000}:

\begin{equation}
K\left(\bm{\xi_{i}}\rightarrow \bm{\xi_{r}}\right)\ge0
\label{CL_non_negativity}
\end{equation}

\noindent derives by the correlation existing between kernels and probabilty density functions. These last, being the derivative of the distribution function, are always non-negative in $\mathbb{R}$.
\noindent The normalisation condition \citep{Cercignani2000}:
\begin{equation}
\int_{\xi_{r,n}}K\left(\bm{\xi_{i}}\rightarrow\bm{\xi_{r}}\right)d\bm \xi_{\bm r}=1
\label{normalizzazione}
\end{equation}

\noindent results from imposing balance between the mass flow arriving at the surface and the mass flow leaving the surface (see Eq. \ref{balance_at_the_surface}). From a physical point of view, normalising the scattering kernels is equivalent to stating that the particles involved in the interaction leave the boundary, so that no capture occurs.
\noindent Finally, the reciprocity or detailed balance condition \citep{Cercignani2000}:

\begin{equation}
f_{0}\left(\bm{\xi_{i}}\right)\lvert\xi_{i,n}\rvert K\left(\bm{\xi_{i}}\rightarrow\bm{\xi_{r}}\right) = f_{0}\left(\bm{\xi_{r}}\right)\lvert \xi_{r,n}\rvert K\left(-\bm{\xi_{r}}\rightarrow -\bm{\xi_{i}}\right)
\label{reciprocity}
\end{equation}

\noindent is a balance equation according to which, for a gas-surface system in thermodynamic equilibrium, a correspondence can be established between 1) the scattering path of the particles and 2) the path that those particles would hypothetically follow if they travelled with velocities that are opposite to those considered in the same time interval \citep{Cercignani1990}.
 This constraint stems by writing the boundary condition as a function of the temperature of the wall. In Eq. \ref{reciprocity}, $f_{0}$ indicates the Maxwellian velocity distribution function corresponding to $T_{w}$.

\subsection{The Maxwell Model}
\label{The Maxwell Model}
\begin{figure*}[bt]
\centering
\def\svgwidth{100mm}
\includegraphics[width = 100mm]{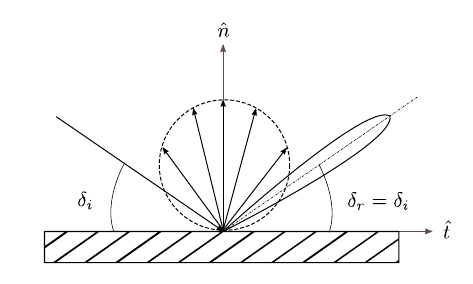}
\caption{Schematic representation of the polar plot distribution predicted by the Maxwell model \citep{Maxwell1890}.}
\label{Maxwell}
\end{figure*}

In Maxwell's model \citep{JClerkMaxwell2015} the behaviour of the reflecting surface is described by the linear combination of the two classical scattering characteristics mentioned in the previous paragraph. According to this, a fraction of incident particles, identifiable with the accommodation coefficient $\sigma$,  is assumed to be trapped by a perfectly absorbing wall. After achieving complete accommodation, the particles are re-emitted from the surface with a Maxwellian velocity distribution typical of a gas at rest and in thermal equilibrium with the wall ($T_{g}=T_{w}$). The remaining fraction $\bigl(1-\sigma\bigr)$ of the incident gas particles collides on an ideally smooth and elastic surface so that, after the collision, the momentum exchange occurs just along the normal direction and the reflected gas lies along the specular direction.
The scattering kernel for this model is accordingly built as the sum of the scattering kernels for specular reflection and diffuse re-emission with complete thermal accommodation \citep{Padilla2009, Dadzie2004}:

\begin{equation}
K_{M}\bigl(\bm{\xi_{i}}\rightarrow\bm{\xi_{r}}\bigr)= \bigl(1-\sigma\bigr)\delta\bigl(\bm{\xi_{i}}-\bm{\xi_{r,specular}}\bigr)+\sigma f_{0}\bigl(\bm{\xi_{r}}\bigr)\lvert\xi_{r,n}\rvert
\end{equation}

\noindent where $\delta$ is the Dirac delta function. This isotropic scattering kernel satisfies the physical constraints discussed and, because of its simplicity, has found substantial success in DSMC implementations. According to the model however, scattering cosine distributions presenting peaks in proximity of the the specular direction (Figure \ref{Maxwell}) should be expected. As discussed in Section \ref{gas_beam}, experimental results show a more complex behaviour which is not predictable by the linear combination adopted by Maxwell, thus limiting the range of applicability of the model.

\subsection{The Nocilla-Hurlbut-Sherman (NHS) Model}
\label{The Nocilla-Hurlbut-Sherman (NHS) Model}
The NHS model, first proposed by Nocilla \citep{Nocilla1964, Nocilla1967} and refined by Hurlbut and Sherman \citep{Hurlbut1968} postulates a "drifting/shifted Maxwellian" law for the reflected particles distribution:

\begin{equation}
f_{r}\bigl(\bm{\xi}\bigr)= \frac{n_{r}}{\bigl(2\pi RT_{r}\bigr)^{3/2}}\exp\Bigg[-\frac{\bigl(\bm{\xi}-\bm{v_{r}}\bigr)^{2}}{2RT_{r}}\Biggr]
\label{drift Max reflection}
\end{equation}

\noindent where $n_{r}$ is the scattering molecules density, $T_{r}$ is the temperature associated with the scattered distribution, and $\bm{v_{r}}$ is the macroscopic or bulk velocity (drift) of the reflected particles. According to Equation \ref{drift Max reflection}, the reflected flux is described by $v_{r}$, the scattering angle $\delta_{r}=90^{\circ} -\theta_{r}$ determined by the direction of $\bm{v_{r}}$ with regards to the tangent to the surface, and $T_{r}$ which in general can be different from $T_{s}$.
In the original form proposed by Nocilla, the model contains the complementary cases of diffuse scattering ($v_{r}=0$) and specular reflection ($s_{r}=s_{i}$, $\bm{v_{r}}=\bm{ v_{i}}$, $\delta_{r}=\delta_{i}$).
However, the mathematical formulation inherently suffers from the lack of connection between the incident and reflected velocity distributions. In this regard, Hurlbut and Sherman \citep{Hurlbut1968} introduced a drifting Maxwellian velocity distribution for the incident particles:

\begin{equation}
f_{i}\bigl(\bm{\xi}\bigr)= \frac{n_{i}}{\bigl(2\pi R T_{i}\bigr)^{3/2}}\exp\Biggl[-\frac{\bigl(\bm{\xi}-\bm{v_{i}}\bigr)^{2}}{2RT_{i}}\Biggr] 
\label{incident distribution NHS}
\end{equation}

\noindent and related the two distributions by determining $n_{r}$ from the equivalence of the incoming and reflected number fluxes in stationary conditions.
%
%
%
A model reformulation was also proposed to reduce the difficulties encountered in matching the experimental results with the model, which requires an accurate determination of $T_{r}$. A partial thermal accommodation coefficient $\alpha_{T,P}\bigl(\theta_{i}\bigr)\sim\alpha_{T,P}\bigl(\delta_{i}\bigr)$, avaraged  over all the possible scattering directions and defined for a specific angle of incidence $\delta_{i}$, was accordingly introduced \citep{Hurlbut1968}:
\begin{equation}
\alpha_{T,P}\bigl(\delta_{i}\bigr)=\frac{E_{i}\bigl(\delta_{i}\bigr)-E_{r}}{E_{i}\bigl(\delta_{i}\bigr)-E_{w}}
\end{equation}
and analytic expressions for the computation of the aerodynamic drag and lift coefficients were determined. For a given angle of incidence of the flow, the aerodynamic coefficients are built as a function of $v_{r}$ (or $s_{r}$), $\delta_{r}$ and $\alpha_{T,P}\bigl(\delta_{i}\bigr)$. Cercignani and Lampis \citep {Cercignani1995} provided a reformulation of the NHS model in the context of the scattering kernel theory, following some suggestions already present in Nocilla's original proposal. Corrections to this model were also applied so that the proposed kernel could satisfy the normalisation (Eq. \ref{normalizzazione}) and detailed balance conditions (Eq. \ref{reciprocity}) \citep{Cercignani1997}. Despite these improvements and some early applications of the model for the computation of the aerodynamic coefficients in FMF conditions \citep{Collins1994}, further implementations did not find much success. However, the good agreement achieved with some gas-beam experimental results \citep{Hurlbut1968} for clean surfaces made the NHS model a fundamental starting point for advances in the study of gas-surface interaction. 

\subsection{The Cercignani-Lampis-Lord (CLL) Model}
\label{The Cercignani-Lampis-Lord (CLL) Model}
The Cercignani-Lampis \citep{Cercignani1971} model (CL) is one of the most successful kernel-based representations of the gas particles-surface interaction as it is described by experimental results. The expression for the scattering kernel is obtained assuming no adsorption and independent interaction of each gas particle with the surface \citep{Cercignani1971}:


\begin{equation}
\begin{split}
\label{Kernel_CL}
K_{CL}\bigl(\bm{\xi_{i}}\rightarrow\bm{\xi_{r}}\bigr) =&\frac{1}{\alpha_{n}\sigma_{t}\bigl(2-\sigma_{t}\bigr)}\exp\biggl[\frac{\alpha_{n}-1}{\alpha_{n}}\bigl(\xi_{r,n}^{2}+\xi_{i,n}^{2}\bigr) +\\& -\frac{\bigl(1-\sigma_{t}\bigr)^{2}}{\sigma_{t}\bigl(2-\sigma_{t}\bigr)}\bigl(\xi_{r,t}^{2}+\xi_{i,t}^{2}\bigr)+\\&+\frac{2\bigl(1-\sigma_{t}\bigr)}{\sigma_{t}\bigl(2-\sigma_{t}\bigr)}\bigl(\bm{\xi_{i,t}}\cdot\bm{\xi_{r,t}}\bigr)\biggr]I_{0}\biggl[\frac{2\bigl(1-\alpha_{n}\bigr)^{1/2}}{\alpha_{n}}\xi_{r,n}\xi_{i,n}\biggr]
\end{split}
\end{equation}

\noindent where for both the incident and reflected velocity, $\xi_{j,t}= \sqrt{\xi_{j,t'}^{2}+\xi_{j,t''}^{2}}$ and $I_{0}$ is the modified  Bessel Function of the first kind and zeroth order. The density of probability described by Eq. \ref{Kernel_CL} is given by the contributions coming from the variation of the three components of the velocity vector. Since the tangential and the normal components can be treated separately in the model, their individual scattering kernels can accordingly be derived. Following the formulation proposed by \citet{Lord1995} for isotropic surfaces, the scattering kernel for the normal component of velocity can be written: 

\begin{equation}
\begin{split}
K_{CLL}\bigl({{\xi_{i,n}}}\rightarrow{{\xi_{r,n}}}\bigr)=&\frac{2{\xi_{r,n}}}{\alpha_{n}}I_{0}\frac{2\bigl(1-\alpha_{n}\bigr)^{1/2}{\xi_{r,n}}{\xi_{i,n}}}{\alpha_{n}}\times\\&\exp\biggl[-\frac{{\xi_{r,n}}^{2}+\bigl(1-\alpha_{n}\bigr){\xi_{i,n}^{2}}}{\alpha_{n}}\biggr]
\end{split}
\end{equation}

\noindent for which the reciprocity and normalisation conditions take the form: 

\begin{equation}
|\xi_{i,n}|\exp\bigl[-{\xi_{i,n}}^{2}\bigr]K_{CLL}\bigl({\xi_{i,n}}\rightarrow {\xi_{r,n}}\bigr)={\xi_{r,n}}\exp\bigl[-{\xi_{r,n}}^{2}\bigr]K_{CLL}\bigl(-{\xi_{r,n}}\rightarrow -{\xi_{i,n}} \bigr)
\label{n_reciprocity_Lord}
\end{equation}

\begin{equation}
\label{n_normalization_Lord}
\int_{0}^{\infty}K_{CLL}\bigl({\xi_{i,n}}\rightarrow{\xi_{r,n}}\bigr) d\xi_{r,n}=1
\end{equation} 

\noindent The interaction phenomenon in the tangential directions is described by the same accommodation coefficient so that, following surface isotropy, the expressions of the scattering kernels for the two tangential components of velocity are in the form of:

\begin{equation}
K_{CLL}\bigl({\xi_{i,t}}\rightarrow {\xi_{r,t}}\bigr)=\frac{1}{\sqrt{\pi\sigma_{t}\bigl(2-\sigma_{t}\bigr)}}\times\exp\biggl\{-\frac{\bigl[{\xi_{r,t}}-\bigl(1-\sigma_{t}\bigr){\xi_{i,t}}\bigr]^{2}}{\sigma_{t}\bigl(2-\sigma_{t}\bigr)}\biggr\}
\end{equation}

\noindent which satisfies:

\begin{equation}
\exp\bigl[-{\xi_{i,t}}^{2}\bigr]K_{CLL}\bigl({ \xi_{i,t}}\rightarrow  {\xi_{r,t}}\bigr)=\exp\bigl[-{\xi_{r,t}}^{2}\bigr]K_{CLL}\bigl(-{\xi_{r,t}}\rightarrow-{\xi_{i,t}}\bigr)
\end{equation}

\begin{equation}
\int_{-\infty}^{\infty}K_{CLL}\bigl({\xi_{i,t}}\rightarrow {\xi_{r,t}}\bigr)d\xi_{r,t}=1
\end{equation}

\noindent As the equations above suggest, the dynamic of the interaction is regulated by two adjustable parameters: the normal energy accommodation coefficient ($\alpha_{n}$) and the tangential momentum accommodation coefficient ($\sigma_{t}$). Lord contributed significantly to the success of the CL model whilst adapting it for DSMC implementation \citep{Lord1991a} and further extended its validity to cases excluded by the original model, among which diffuse re-emission with incomplete accommodation \citep{Lord1991, Lord1995} . Because of this, it is generally preferred to refer to the model as the Cercignani-Lampis-Lord model (CLL).

A possible implementation of the CLL model in closed-form solutions was proposed by Walker et al. \citep{Walker2014}. The authors suggested modified expressions of the Schaaf and Chambre \citep{Schaaf1958, Storch2002} (S\&C) closed form equations with the CLL model. 
 The attempt is rather difficult since, among other parameters (Fig. \ref{model_param}), the S\&C closed-form solutions are written as a function of $\sigma_{n}$ and $\sigma_{t}$. While an immediate relation can be found between the tangential momentum accommodation coefficient and the tangential energy accommodation coefficient ($\alpha_{t}$):

\begin{equation}
\alpha_{t}=\sigma_{t}\bigl(2-\sigma_{t}\bigr)
\end{equation}

\noindent the same cannot really be said for $\alpha_{n}$ and $\sigma_{n}$. Approximate analytic $\sigma_{n}-\alpha_{n}$  relations, written as a function of four best-fit parameters, were found adopting a least squares error approach and sensitivity analysis. Ranges of variation were selected for some meaningful parameters about their nominal values. In this way the agreement between the $\sigma_{n}-\alpha_{n}$ relation and the expected $C_{D}$ values could be evaluated for the nominal conditions and over the range of variation of the selected parameters. These last were identified with the bulk velocity of the particles, the free stream temperature, the surface temperature, the normal thermal accommodation coefficient and the tangential momentum accommodation coefficient. 
Good correlation between the computed $C_{D}$ and the values provided by the CLL model implemented in DSMC Analysis Code was seen for the modified S\&C closed-form solutions written as a function of the derived $\sigma_{n}-\alpha_{n}$ laws. However, the set of values to be chosen for the best-fit parameters is not constant but varies with the gas species considered, the type of body impinged and, in the case of lighter molecular species, the $\alpha_{n}$ range assumed. Values suggested by Walker et al. for representetive molecular species and body shapes could be found in the original paper from the authors \citep{Walker2014}. Bigger uncertainties are found in the case of He and H for value of $\alpha_{n}$ close to unity.

%
%
%

\section{Physical GSI Models} \label{Physical GSI Models}
Physical GSI models take advantage of experimental results to describe how the thermal motion of the surface influences the scattering dynamics of the impinging gas-beam. These models are thus based on assumptions regarding the surface interaction potentials, the surface morphological structure and the surface elasticity/stiffness characteristics. Among the vast number of models present in literature, special attention will be devoted to the simple quasi one-dimensional Hard Cube model and its most successful expansions: the Soft Cube model and the Washboard model. Two and three-dimensional lattice models \citep{Oman1968a,  Raff1967, Lorenzen1968, McClure1969} will be neglected in this review. These last are generally characterised by a more complex implementation which leads to higher accuracy and also computational time; factors potentially limiting the range of applicability of these models for the context of this paper. Moreover, if the increase in complexity is justified in the frame of pure gas-surface interaction, the same might not be true for orbital aerodynamics engineering applications. The number and range of uncertainties affecting the problem of aerodynamic forces and torques estimation in VLEO is quite high. Because of this, the increased level of complexity is likely to be unjustifiable against the numerous sources of errors observed.

\subsection{The Hard Cube Model (HC)}
\begin{figure*}[bt]
\centering
\def\svgwidth{100mm}
\includegraphics[width = 100mm]{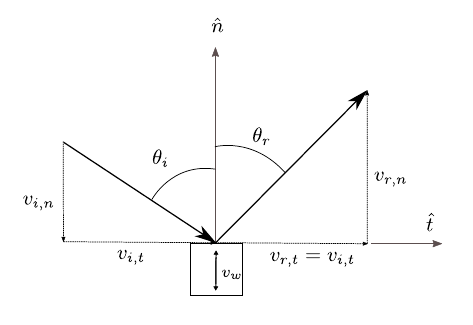}
\caption{Hard Cube Model, reproduced from \citep{Logan1966}.}\label{HardCube}
\end{figure*}

The Hard Cube Model, as proposed in its earlier form by Goodman \citep{Goodman1965}, has found success with Logan's and Stickney's \citep{Logan1966} formulation. Despite its inherent simplicity, resulting from the assumptions adopted in its development, the model is able to qualitatively reproduce the experimental lobal scattering typical of clean and polished surfaces. 
The model assumes the surface to be perfectly smooth and the gas particles and surface atoms involved in the interaction to be ideally elastic and rigid. The dynamics of the collision is simplified  by assuming that each gas particle interacts solely with
a surface atom represented as an isolated cube in the lattice, so that any impact of the surface structure on the scattering properties is neglected. During the collision, the gas particle and the surface atom interact as free particles so that a one-dimensional impulsive-repulsive potential, with no attractive well, can be conjectured. In this way, the impact of interaction times on the collision mechanism can be ignored with benefits in terms of simplicity and with only a partial loss of accuracy. The cubes comprising the surface are oriented so that one of their four faces lies in the direction parallel to the surface contour and they are characterised by an initial Maxwellian normal velocity distribution determined by the surface temperature. The momentum exchange is assumed to be due uniquely to the normal component of the gas particle velocity $\left(v_{n}\right)$ as the tangential component $\left(v_{t}\right)$ is preserved by the surface properties (\figurename~\ref{HardCube}). The model predicts a quasi-specular re-emission both above and below the specular range with each scattering angle $\theta_r$ being determined by a unique value of $v_{r,n}$ for a given $v_{i,t}=v_{r,t}$. 

Along with the numerical formulation, an analytical approach in which mean velocity values are adopted instead of velocity distributions for both the gas particles and the surface atoms was proposed by \citet{Logan1966}. According to this approach, closed form solutions can be derived and the parameters on which the interaction depends can be more easily identified. The flat surface assumption allows the restriction of the analysis to the plane identified by the surface normal and the incident velocity vector, so that:

%
%

\begin{equation}
\bm{\theta_{r}}=\cot^{-1}\left[\cot\left(\theta_{i}\right)\left(\frac{1-\mu}{1+\mu}+\frac{16\mu}{9\pi \left(1+\mu \right)}\frac{T_{w}}{T_{g}\cos^{2}\left(\theta_{i}\right)}\right)\right]
\label{thetas_HC}
\end{equation}

\noindent where $\mu$, the gas particle-surface atom mass ratio, is restricted to vary in the following range: 

\begin{equation}
0<\mu=\frac{m_{g}}{m_{s}}<\frac{1}{3}
\label{mass_ratio_HB}
\end{equation}

\noindent and $T_{g}$ is the gas-beam source temperature. Constraints on the possible values of $\mu$ result from further assuming that the gas particle experiences a single interaction with the surface atom considered.
According to equation \eqref{thetas_HC}, the scattering direction depends on the mass ratio $\mu$, the incidence angle and the surface-to-gas temperature ratio. Simplicity and ease in implementation are however obtained at the price of a general loss in accuracy in describing the experimental results compared to the extended model described in \citep{Logan1966}, for which numerical simulation is needed.

Some expansions of the HC model have been proposed, the most successful one discussed later in this section. Hurst et al. \citep{Hurst1982} and Nichols et al. \citep{Nichols1975, Nichols1975a} modified the model to capture rotational dynamics of elliptically-shaped diatomic molecules scattered from the surface. Doll \citep{Doll1973} addressed the rotational dynamics modelling diatomic homonuclear molecules as rigid rotors with motion restricted to a single plane. The importance for this model of multiple collisions with the surface atom arising from the rotational state were also discussed. Trapping phenomena were addressed by Weinberg and Merrill \citep{Weinberg2002}. Trilling and Hurkmans \citep{trilling1976} introduced an attractive long-range Coulomb potential, an exponential short-range repulsive potential and modified the surface geometry treating atoms as "spherical caps" rather then cubes. Sitz et al. \citep{Sitz1988} expanded the HC model to qualitatively describe momentum orientation in the scatterring of N\textsubscript{2} from smooth Ag(111), introducing a frictional force along the surface tangential direction. The additional level of complexity characterising these models seems inappropriate for the applications for which this review paper is intended and, because of this, they will not be discussed more thoroughly in the following sections.

\subsection{The Soft Cube Model}
\begin{figure*}[bt]
\centering
\def\svgwidth{80mm}
\includegraphics[width = 80mm]{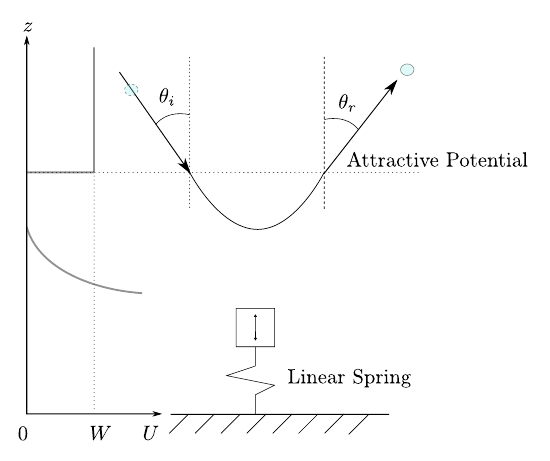}
\caption{The Soft Cube Model, reproduced from \citep{Logan1968}.}\label{SoftCube}
\end{figure*}

The Soft Cube Model proposed by Logan and Keck in 1968 \citep{Logan1968} owes its name to the introduction of a more realistic "soft" potential to capture the physics of the gas-surface interaction. The atoms that comprise the flat surface and take part in the collision are assumed to behave like independent cubes linked to the underlying lattice by means of single linear springs (\figurename~\ref{SoftCube}). Surface atoms are thus regarded as oscillators characterised by a natural frequency $\omega$ and a Maxwellian energy distribution corresponding to the surface temperature $T_{w}$. Similarly to the HC model, the interaction with a gas particle involves a single cube in the lattice and the energy exchange, responsible for the accommodation coefficient value, is due solely to the  variation of the normal component of velocity after the collision $ (v_{i,t}=v_{r,t})$. The interaction is however more realistically captured assuming non-negligible collision times, described by a non-impulsive interaction potential consisting of two components. In addition to a repulsive exponential potential, which substitutes the impulsive repulsive interaction assumed in the HC model, an attractive long-range square-well potential component is introduced. The model can be eventually employed to obtain an estimate of the fraction of particles that remain trapped on the surface after the collision and depart from it after achieving sufficient energy.

%
%
%
%
%
%

Comparison with experimental results is obtained by properly selecting the value of three modifiable parameters: 1) the potential well-depth $W$, 2) the interaction range $b$ and 3) the oscillator frequency $\omega$, whose value is assumed to be given by the Debye temperature $\left(\Theta_{D}\right)$. Combinations of $b$ and $W$ that reproduce, with satisfactory agreement, the experimental data referred to a selection of gas-surface systems can be found in the original paper by \citet{Logan1968}.

%
%

\subsection{The Washboard Model}
Like the Soft Cube Model, the Washboard model \citep{Tully1990} can be regarded as an attempt to improve the Hard Cube model agreement with the experimental results. Compared to the other GSI models discussed so far, the Washboard model has the advantage of addressing the effect of surface corrugation on the scattering properties while preserving relative clarity and simplicity. 

\begin{figure*}[bt]
\centering
\def\svgwidth{130mm}
\includegraphics[width = 130mm]{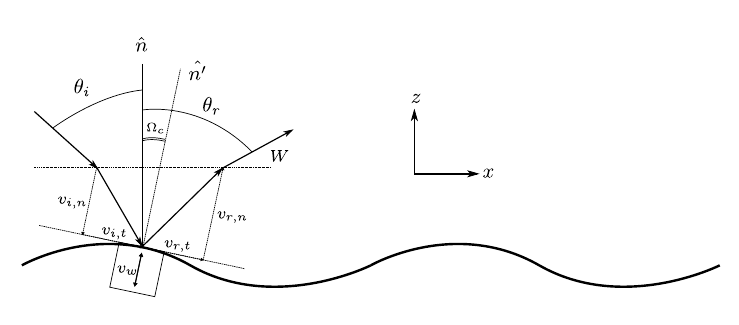}
\caption{The washboard model, reproduced from \citep{Tully1990}.}\label{washboard}
\end{figure*}

The surface contour is simplified assuming a sinusoidal profile applied exclusively in one direction, thus making the model appropriate for bidimensional but not for out-of-plane scattering evaluation. Similarly to the HC model, the cube with which the colliding gas particle interacts is oriented along the surface contour and its velocity is determined by a Maxwellian distribution at the surface temperature. Because of the surface corrugation, the cubes are tilted with regards to the normal to the flat surface so that, for any impact point, a local normal and tangential directions can be identified (\figurename~\ref{washboard}). The maximum deviation of the local normal from the flat surface normal direction is measured by the corrugation strength parameter $\left(\Omega_{C}\right)$.
The introduction of this parameter allows the model to adapt to different levels of surface corrugation, thus providing good qualitative agreement with experimental results ranging from smooth to rough surfaces. 
%
The attractive potential well $W$ produces refraction in the gas particle initial trajectory, thus varying its normal and tangential component of velocity. Even in this case the nature of the interaction is impulsive so that in the local normal-tangential reference system, the tangential momentum is unchanged and the energy exchange is determined solely by the normal momentum variation. As a consquence, in the $xz$ flat surface reference system the tangential momentum is not conserved and the strict assumption characterising both the Hard and the Soft Cube model is accordingly removed. 

Analytic expressions for the angular scattering velocity and kinetic energy distributions were provided for small surface corrugations. The parameters on which these depend are the corrugation strength coefficient $\Omega_{C}$, the incident angle $\theta_{i}$, the incident kinetic energy $E_{i}$, the mass ratio $\mu$, the potential well $W$ and the surface temperature $T_{w}$. Eventually, the trapping probability can also be addressed. Further extensions of the washboard model include the works of Yan et al. \citep{Yan2004} and Liang et al. \citep{Liang2018}.

\section{Comparison of GSI Models with Gas-Beam Experiments}\label{gas_beam}
Molecular beam experiments are a useful means of gaining information concerning the energy exchange at the surface and the wall-gas system characteristics at atomic scale. For appropriate selection of incident energy distributions, the scattering of neutral molecules on a target is representative of orbital conditions and can thus be used to investigate the aerodynamic behaviour of specific materials in the context of space applications. The literature covering the topic is vast and this section does not claim to be exhaustive, since such an activity would require a specific review effort. The objective is thus to provide an overall picture of the subject, focusing the attention on the physical regimes that are significant for the purpose of this paper, thus addressing, wherever possible, the points of strength and the limits of the models previously discussed in the context of VLEO aerodynamics. At VLEO altitudes, AO is the dominant constituent of the residual atmosphere, with atoms impacting on the exposed surfaces with an average velocity of $\sim 7.8\: km\:s^{-1}$ corresponding to a Maxwellian mean incident energy distribution of $\sim5\:eV$. According to this, attention will be devoted especially to results referring to molecular beam scattering of monoatomic species from targets in a variety of conditions. Studies analysing diatomic and polyatomic beams scattering are numerous but their results are typically more difficult to interpret: the interaction depends on both the translational and internal energies of the molecule considered and on the local aspect of the interaction potential. However, results for N\textsubscript{2} and O\textsubscript{2} scattering from Ag(111) reported by Asada et al. \citep{Asada1982} show mean velocity and mean energy distributions with scattering angle which are similar to those obtained for monoatomic molecules. Similarly, analysis of adsorption and desorption rates, which are important especially for heavier molecular species, requires correlation between the characteristics of the system interaction potential and the sticking probability. These, in turn, vary with initial rotational and translational energies, binding energies, orientation of the molecules, incidence angle and surface temperature \citep{Muhlhausen1985}. Because of this, the spatial distributions obtained are the result of a complex mechanism of interaction \citep{Palmer1970,Smith1964a} and more advanced numerical tecnhiques involving molecular dynamics or binary collision approximation are required. Extensive reviews on the topic are however available in \citep{Goodman1971,Barker1985,Bernasek,Somorjai1973,Cardillo2003a, Weinberg1975} with some more dated results reported in \citep{Hurlbut1962}.

The scattering behaviours observed are not constant as they tend to change substantially with the system considered and, in particular, with the ratio between the mass of the gas particles and the surface atoms, the range of interaction, the energy/temperature of the incident beam with regards to the surface temperature, the molecular or atomic species involved in the experiment, the presence of adsorbents on the target surface, the morphology of the surface considered despite the level of roughness and the relative position and orientation of the gas particles and surface atoms \citep{Goodman1971}. At the time of writing, a comprehensive theory capable of capturing each possible scenario is not available such that multiple models, suitable for specific physical regimes, are adopted instead. 
For light particles and low incidence energies \citep{Barker1985} compared to solid maximum phonon energy, the interaction is predominatly elastic. In these conditions, no energy transfer occurs in the system and quantum mechanical phenomena, such as diffraction, are expected to be predominant \citep{Goodman1971}.

 As gas particles mass and incident energy increases, the collisions become more inelastic in nature and classic theory is adequate enough to reproduce the scattering behaviour. For this physical range, whether particles gain or transfer energy to the surface depends on the relative magnitude of $E_{i}$ and $T_{w}$. Generally three mechanisms of interaction are identified: 1) Single gas-surface collision with moderate net energy exchange; 2) Multiple gas-surface collisions with no adsorption and delayed scattering; 3) Multiple gas-surface collisions with adsorption to the surface and eventual desorption in the scattered gas. The latter interaction mode is typical of highly contaminated surfaces \citep{Hurlbut1957, Bishara1970,Yamamoto1970a}: in these conditions, the adsorbed particles have time to reach equilibrium with the surface and they are scattered according to a cosine distribution with $\theta_{r}$ and a Maxwellian translational velocity distribution corresponding to the wall temperature. The majority of works published in literature, however, refer to the first two interaction scenarios, typically observed in scattering from clean flat surfaces in Ultra High Vacuum (UHV) conditions. In this case, lobal re-emission distributions characterised by a predominant asymmetrical quasi-specular component are observed for varying surface properties, incident particles energies and wall temperature \citep{Asada1982, Murray2017, Mehta2018, Yamamoto1970a, Moran1969, Smith1964, West1971, Smith1964}. The generally small amount of particles that undergo multiple collisions before being scattered in the gas-phase determines the width of the lobe of distribution and seems to be susceptible to the morphology of the surface considered, despite the level of roughness. Wider angular distributions were obtained for smoother surfaces even when higher incident energies were employed \citep{Murray2017}. This seems to suggest that the interactions within atoms in the same surface layer or adjacent layers may play a role in determining the scattering characteristic.      

\begin{figure*}[h]
\centering
\def\svgwidth{120mm}
\includegraphics[width = 120mm]{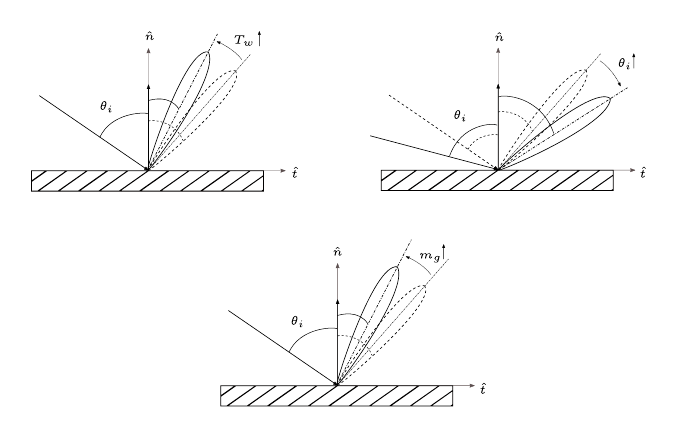}
\caption{Variation of $\theta_{r,max}$ - the angle at which the peak of the distribution occurs - in the thermal scattering regime. Behaviours observed with increasing $T_{w}$ (top left), $\theta_{i}$ (top right) and $m_{g}$ (bottom) are illustrated.} 
\label{ThermalCharacteristics}
\end{figure*}

In the inelastic scattering domain, however, different scattering behaviours and thus different trends are expected according to the incident energy and the interaction radius for the system considered \citep{Oman1968}. As highlighted by Goodman in \citep{Goodman1971}, for gas-surface systems that are not characteristed by a strong periodicity in the interaction potential,  low values of incident kinetic energy ($E_{i}<k_{B}T_{w}$) and large interaction distances \footnote{The interaction distance is defined here following the definition of the non-dimensional radius parameter \textsl{R} provided by Goodman in \citep{Goodman1971}. The distance of interaction is thus defined as the ratio between two quantities: the closest distance that separates the centre of the impinging atom from the centre of the surface atom during the collision and the critical value of this distance for the gas-surface system considered. When the critical distance value is achieved the impinging particle enters the surface. } define the so-called \textit{thermal scattering} regime. In these conditions, the impinging gas-particles can not see the surface corrugation and the surface appears flat and smooth. During the interaction, the tangential component of momentum is generally conserved and the scattering dynamics are dominated by the surface thermal motion in the direction normal to the surface.  Thermal scattering studies \citep{Yamamoto1970a, Moran1969, Stoll1971, Fisher1968, Romney1969, Sau1973} show the following common features for the angular distributions (Fig. \ref{ThermalCharacteristics}):

\begin{enumerate}
\item{$\partial\theta_{r,max}/\partial T_{w}\le 0$}: the scattering angle corresponding to the peak in the distribution slightly moves towards the normal to the surface with increasing $T_{w}$ \citep{Rettner1991}. Moreover, at high values of $T_{w}$, for which the surface appears to be free of adsorbents, the width of the distribution experiences a slight increase with decreasing $T_{i}/T_{w}$ \citep{Yamamoto1970a, Stoll1971}. Higher $T_{w}$ also induces lower scattering intensity at the peak of the distribution \cite{Stoll1971, Sau1973}. Some authors, however, obtained the opposite trend for the scattering of Ar \citep{Weinberg1972}, Xe \citep{Weinberg1972, Sau1973} and Kr \citep{Stoll1971, Weinberg1972}  on various metal surfaces at $T_{i}=T_{amb}$. Generally speaking, high wall temperatures are efficient in reducing the sticking probability and the time required for the interaction, thus preventing complete accommodation;
\item{$\partial\theta_{r,max}/\partial\theta_{i}\ge 0$}: the scattering angle corresponding to the peak in the distribution moves towards the surface tangent as the incidence angle increases \citep{Logan1966, Yamamoto1970a, Fisher1968, Romney1969}. Moreover, {$\partial\theta_{r,max}/\partial m_{g}\le 0$}, i.e. the scattering angle for which the distribution presents a peak moves towards the normal to the surface as the mass of the incident gas atom increases \citep{Saltsburg1966, Yamamoto1970a, Logan1966};
\end{enumerate}

\begin{figure*}[bt]
\centering
\def\svgwidth{100mm}
\includegraphics[width = 100mm]{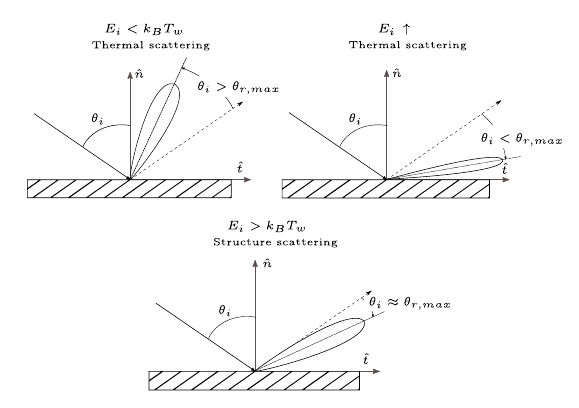}
\caption{Transition from thermal to structure scattering with increasing values of  $E_{i}$.} 
\label{ThermalStructure}
\end{figure*}

As the incident kinetic energy of the beam increases (within the limits of the thermal scattering domain), the effect of surface thermal vibration on the angular scattering distribution becomes less dominant. As a consequence, the lobal distribution becomes narrower and more symmetrical in shape, the scattered intensity at the peak increases \citep{Stoll1971} and $\theta_{r,max}$ generally moves towards the tangent to the surface \citep{Oman1968, Mehta2018, Saltsburg1966, Berenbak2002} (\figurename~\ref{ThermalStructure}, top right). When this condition is observed, the re-emission distribution is said to be \textit{superspecular}. The effect seems to be more noticeable for nearly grazing angles of incidence \citep{Mehta2018}. With regards to translational energy distributions, when the conservation of the tangential momentum is observed, the relative ratio between the mean final and incident energies varies according to the parallel momentum conservation curve:

\begin{equation}\label{PMC}
\frac{E_{r}}{E_{i}}= \frac{\sin^{2}\theta_{i}}{\sin^{2}\theta_{r}}
\end{equation}

\noindent The characteristics mentioned above are well described by the cube models \citep{Rettner1991} reviewed in the previous section of this paper, because of the inherent assumptions on which these models are built. Better agreement, as expected, is found with the SC \citep{Bishara1970, Yamamoto1970a, Logan1968} rather than with the HC model \citep{Alexander2008, Mehta2018, Oman1968, Rettner1991, Sau1973, Stoll1971}, not only because of the more realistic gas-surface interaction potential assumed, but also because adjustments can be made through the parameter $W$ for the system considered. When the attractive well $W$ dominates the dynamics, the repulsive potential assumption loses accuracy and the HC model fails in describing trapping and partial accommodation to the surface. The limitations imposed on the gas particle-surface atom mass ratio in the HC model are likely to make the model unsuitable for addressing atomic oxygen scattering from most surfaces. Moreover, comparison of the potentialities of these two models against experimental data is possible just in the incident scattering plane. For out of plane scattering considerations, techniques addressing surface corrugation in more than one dimension need to be employed. While Maxwell's model fails in reproducing the petal-shaped angular distribution observed in these experiments, it appears that its theoretical apparatus and simplified assumptions are sufficient to describe some re-emission polar plots showing a small nearly specular and a large diffuse component \citep{Murray2017, Hurlbut1957}. Results provided by Mehta et al. \citep{Mehta2018} show that when the CLL model is adopted some difficulties are encountered in the attempt of selecting the proper combination of $\alpha_{n}$ and $\sigma_{t}$ to reproduce the experimental conditions. Excellent predictions of the position of the peak ($\theta_{r,max}$) and of the dispersion of the scattering distribution, for selected values of the accommodation coefficients, seem to exclude an accurate representation of the experimental $E_{r}/E_{i}$, and viceversa.

As the kinetic incident energy of the beam increases \citep{Miller1970, Rettner1991, Wiskerke1992} with regards to the surface atoms thermal energy ($E_{i}>k_{B}T_{w}$) and the radius of interaction reduces, transition to the \textit{structure scattering regime} is experienced: in this case, the surface roughness sensed by the impinging particles is noticeable because of the increased power of penetration. The interaction is no longer dominated by the surface thermal motion but by the surface corrugation. Because of the multiple collissions experienced by the particles with the rough surface, the angular distribution in this regime becomes wider in shape, the value of the peak scattered intensity decreases and a shift from superspecular to directions closer to the specular range for $\theta_{r,max}$ is observed \citep{Romney1969,Miller1970a}. The qualitative re-emission behaviour expected for increasing value of $E_{i}$ in the transition from the thermal to the structure regime are reproduced in \figurename~\ref{ThermalStructure}. In contrast to the thermal regime, the energy of the scattered beam ($E_{r}$) increases with increasing $\theta_{r}$ \citep{Rettner1991, Zhang2002}.  Cube models relying on the flat surface approximation, are unable to reproduce this scenario and agreement is rather found with the Hard Spheres model \citep{Goodman1967}, the Washboard model \citep{Berenbak2002}, and, in  general, more complex models. Accordance of the HC model with some gas-beam data referred to hyperthermal $E_{i}$ seems fortuitous \citep{Watanabe2006} and, according to the authors, attributable to the morphology of the gas-surface system considered. An interesting scattering phenomenon observed in the structure regime is rainbow scattering: when this process occurs the typically wide spatial distribution is characterised by two peak lobes corresponding to different intensities in the flux measurements. This phenomenon was however, predominatly observed on the scattering of rare gases from LiF surfaces \citep{Smith1970a,Tomii2000}, with some evidence on systems composed by metal surfaces \citep{Berenbak2002, Amirav1987}. More details on the topic can be found in \citep{Goodman1971, Barker1985}. The appearance of rainbow scattering effects seems to be associated with surface corrugation. In this regard, the Washboard model offers a relatively simple formulation able of capturing more complicated re-emission mechanisms for which the HC and SC models are not suited. Moreover, the Washboard model appears to be more effective in describing gas particle attraction and surface penetration as well as scattering characteristic from softer surfaces \citep{Alexander2008}, especially if compared with the HC model. These features are particularly useful when studying gas-surface interaction specifically for orbital aerodynamic applications. The interaction of most materials with the orbital environment is likely to cause variations in the static gas-surface potential corrugation \citep{Banks2004} and, more in general, degradation of material performances with time. The lattice structure is subjected to become rougher as AO exerts its erosion action on the ram surfaces. As a consequence, materials with promising aerodynamic performances (i.e. near-specular re-emissions) might experience variations in the expected scattering behaviour within a mission lifetime. In this scenario, the principles on which the Washboard model is built could prove useful to address materials degradation and performance variation.

\section{Conclusions}\label{Conclusions}
 Renewed interest in small satellites missions in the lower region of the Earth's atmosphere demands new aerodynamic technologies capable of taking advantage of the environment in LEO. Aerodynamic performances are dependent on the mechanisms that rule the gas-solid interaction, but great uncertainties are associated with the physical processes occurring at the wall in rarefied and extremely rarefied regimes. Gas-beam experimental results suggest that reality might be more complex than that described by classical theoretical kinetic models. The development of a new generation of aerodynamic materials may therefore require more accurate predictions. The number of uncertainties in the system behaviour reflects a vast production of models in literature: difficulties however arise in the attempt of combining efficiency with simplicity. In the previous sections popular models which have obtained considerable success for their immediacy and ability to predict scattering behaviours have been reviewed. These models were broadly associated with two principal families according to some common features. Scattering-kernel theory based GSI models are built on a statistical approach, while physical GSI models  provide a simple tools to describe the complex physical interaction mechanisms observed at the wall. Wherever possible, their appropriateness has been discussed against relevant gas-beam experimental results for the problem considered and physical ranges of application have been identified.
 
The joined effort of several authors has resulted in remarkable improvements in the understanding of the phenomena involved in the observed re-emissions. The problem is however complex and multidisciplinary in nature. Despite the challenges that remain, they represent a starting point for future developments. At the moment of writing, it appears that an easy-to-implement model applicable to different scattering regimes and to different gas-solid systems is still to be defined. Similarly, a simple analytical model capable of a more accurate quantitative description of the behaviours of both clean and contaminated surfaces might be desirable. In this regard, the level of technological advancement achieved by gas-beam facilities seems adequate to support a more critical analysis of the models developed so far. A more scrupulous examination of the approximations on which these rely may help in identifying the points of strength of each model and possibly expand their range of applicability. There might also be a chance to identify with more accuracy the re-emission patterns that an improved GSI model should be able to describe. A possible strategy may include enriching the proposed scattering-kernels for GSI with some more realistical assumptions regarding adsorption and desorption phenomena. Further comparison of theoretical models addressing surface corrugation with a broader range of experimental data might be helpful as well. Due to the wide amount of results concerning scattering from clean surfaces, the majority of works seem to focus mainly on comparison with the simpler HC and SC models. The Washboard model appears to be privileged in the study of rainbow scattering from extremely corrugated surfaces. However, using this model against data referring to less corrugated surfaces might facilitate the understanding of the GSI problem.

Some adaptability characteristics are especially desired when VLEO aerodynamic applications are discussed. The growing interest in Earth-observation missions at these altitudes has paved the way for the study of aerodynamic materials promoting nearly specular reflection. However, the results obtained in a controlled facility may be affected by considerable alterations in the real thermospheric environment. A robust GSI model capable of providing satisfactory agreement for a range of interaction performance can facilitate the task of designing aerodynamic features for VLEO satellites. As a consequence, increase in the reliability of the aerodynamic control manoeuvres proposed in literature is expected. End-of-life tasks, satellite geometry and ABEP system design are likely to benefit from any knowledge improvement as well. This task, although difficult, seems to be easier to achieve in the context of orbital aerodynamics. The requirements imposed on the accuracy of the model are more relaxed than those expected in the more general frame of gas-surface interaction science. This is a result of the considerable number of additional uncertainties that affect the estimation of the aerodynamic forces and torques. A less accurate but still effective model may therefore adequately serve the purpose of describing a range of scattering behaviour or, equivalently, aerodynamic performance.  

\section{Acknowledgements}\label{Acknowledgements}
The DISCOVERER project has received funding from the European Union's Horizon 2020 research and innovation programme under grant agreement No 737183.
Disclaimer: This publication reflects only the views of the authors. The European Commission is not liable for any use that may be made of the information contained therein.

\bibliographystyle{elsarticle-num-names}
\bibliography{C:/Users/Sabrina/Documents/BibTeX/library}



%
\end{document}